\documentclass[a4paper]{jpconf}
\usepackage{graphicx}
\usepackage{amsmath,amssymb}
\usepackage{caption,subcaption}
\usepackage{hyperref}

\begin{document}
\title{Supernova Constraints on Dark Flavored Sectors}

\author{Jorge Terol-Calvo}

\address{
    Instituto de Astrof\'isica de Canarias, C/ V\'ia L\'actea, s/n E38205 - La Laguna, Tenerife, Spain\\
    Universidad de La Laguna, Dpto. Astrof\'isica, E-38206 La Laguna, Tenerife, Spain}
\ead{jorgetc@iac.es}

\begin{abstract}
    Proto-neutron stars forming a few seconds after core-collapse supernovae are hot and dense environments where hyperons can be efficiently produced by weak processes. By making use of various state-of-the-art supernova simulations combined with the proper extensions of the equations of state including $\Lambda$ hyperons,
    we calculate the cooling of the star induced by the emission of dark particles $X^0$ through the decay $\Lambda\to n X^0$. 
    Comparing this novel energy-loss process to the neutrino cooling of SN 1987A allows us to set a stringent upper limit on the branching fraction, BR$(\Lambda\to n X^0)\leq 8\times10^{-9}$, that we apply to massless dark photons and axions with flavor-violating couplings to quarks. We find that the new supernova bound can be orders of magnitude stronger than other limits in dark-sector models.   
\end{abstract}

\section{Introduction}
The nature of dark matter remains one of the most important questions in physics. Many possible solutions have been propsed \cite{Bertone:2010zza}, among them, an atractive possiblity is the existence of a dark sector, not coupled to the Standard Model (SM) particles directly but interacting with them through new mediators also known as \emph{portals}. 
Axions and axion-like-particles (ALPs) originating from a spontaneously broken global symmetry or dark photons, gauge bosons of a dark gauge symmetry, are leading exemples of bosonic portals to a dark sector. In principle, these portals could have a rich flavour structure, depending on the ultraviolet (UV) completion of the model. Therefore,
rare meson and lepton decays can be rather large in these models and, thus, the search of these processes in laboratories can set strong constraints on these models.  

Alternatively, dark sectors can also be constrained from energy-loss arguments applied to stellar evolution, typically constraining the coupling of light particles (masses below the temperature of the stellar medium) to photons, nucleons or electrons. However, there is a system where one can prove couplings to heavier flavours of the SM, that is the proto-neutron star (PNS) formed during
core-collpase supernovae (SN). The temperatures and densities reached in the PNS allow for the production of muons \cite{Bollig:2017lki} and lambda hyperons \cite{Oertel:2016bki}.

The core-collapse SN mechanism has been well studied from the observation of SN 1987A with neutrinos and a limit has been set on alternative cooling mechanisms with new particles or interactions, that is, on the \emph{dark luminosity} $L_{\rm d}$,

\begin{equation}
    L_{\rm d}\lesssim3\times10^{52}~~\text{erg s$^{-1}$}, \label{eq:boundRaffelt}  
\end{equation} 
at $\sim1 \,$s after bounce.

\section{Strange cooling mechanism}

Hyperons produced in the PNS can be the source of a new dark cooling mechanism, in particular, the existence of a new dark boson $X^0$ that interacts with strange and down quarks in a way that permits the decay $\Lambda\to n X^0$ would produce this new cooling mechanism. One can compute the spectrum of the dark cooling rate and find

\begin{equation}\label{eq:Nexact}
    \frac{d Q}{d\omega}= \frac{m_\Lambda^2\Gamma \omega}{2\pi^2\bar\omega} \int^\infty_{E_0} dE \, f_\Lambda(1-f_n),
\end{equation}

where $\Gamma\equiv\Gamma(\Lambda\to n X^0)=\frac{\bar\omega^3}{2\pi} C_{X}$ is the width of the decay $\Lambda\to n X^0$ for a massless $X^0$, in vacuum and the $\Lambda$'s rest frame, $\bar\omega=(m_\Lambda^2-m_n^2)/2m_\Lambda$ is the $X^0$ energy in this frame, $m_{\mathfrak B}$ (${\mathfrak B}=n,~\Lambda$) are the baryon masses and $C_{X}$ is a constant with dimensions of $E^{-2}$ that is related to the energy scale and couplings of the model and $\omega$ ($E$) is the energy of the $X^0$ ($\Lambda$) 
in the PNS's rest frame. The number densities of the baryons follow the relativistic Fermi distributions, $f_\mathfrak{B}$, at a given temperature, $T$, and chemical potential, $\mu$, established by ``$\beta$-equilibrium'', $p e^-\leftrightarrow {\mathfrak B}\nu_e$. 

One can get a rather simple expression for Equation \ref{eq:Nexact} by making some approximations. If we go to the nonrelativistic regime, ignore Pauli blocking and treat the species as free Fermi gasses we get the expression
\begin{equation}
    \label{eq:Qapp}
    Q\simeq (m_\Lambda-m_n)\, \Gamma\, n_n e^{-\frac{m_\Lambda-m_n}{T}},  
\end{equation}
which only depends on the number density of the neutrons, $n_n$, the decay width in vacuum, $\Gamma$, and the temperature of the PNS, $T$.

\subsection{Reabsorption and trapping}

The $X^0$ can be reabsorbed by the inverse mechanism by the stellar medium if their mean free path is shorter than the size of te PNS. The absorption rate,
\begin{equation}
     \dfrac{d\mathcal N_{\rm ab}}{d\omega}=\dfrac{1}{\omega}\dfrac{d Q_{ab}}{d\omega},
\end{equation}
can be calculated analogous to the emision rate assuming time-reversal symmetry and thermal equilibrium. Then, from the detailed balance one gets the energy-dependent mean-free path 

\begin{equation}
    \label{eq:free_mean_path}    
    \lambda_\omega^{-1}=\frac{1}{\frac{dn_X}{d\omega}}\frac{d\mathcal N_{\rm ab}}{d\omega}=\frac{m_\Lambda^2 \Gamma }{\bar\omega\omega^2}
    \int^\infty_{E_0} dE(1-f_\Lambda)f_n,
\end{equation}
where $n_X$ is the number density of $X^0$ in the medium. From the mean-free path one can get easily the optical depth $\tau$

\begin{equation}
    \tau(\omega,r)=\int^\infty_r\lambda_\omega(r')^{-1}dr',    
\end{equation}
which describes the exponential damping on the emission of $x^0$. The total dark luminosity of the PNS will be then,

\begin{equation}
    \label{eq:dark_lumi}
    L_{\rm d}=\int d^3\vec{r}\int^\infty_0 d\omega \frac{dQ(r)}{d\omega}e^{-\tau(\omega,r)}.  
\end{equation}

For the strong coupling limit where the mean-free path of the $X^0$ is much shorter than the radius of the PNS, the particle experiences several absorptions and emissions. This is known as the \emph{trapping regime} and the emission of $X^0$ is described by 
black-body radiation from a surface where the optical depth is, averaged over $\omega$, equal to 2/3. There is a maximum radius of emission, $R_{\rm d}$, at density and temperature ($T_{\rm d}$) where 
the $\Lambda$ hyperons are no longer produced in the medium. This sets a minimal emission loss-rate in the trapping regime determined by 
\begin{equation}
    \label{eq:emission_trapping}
    L_{\rm d}^{\rm t}= \frac{\pi^3}{30} g_s R_{\rm d}^2 T_{\rm d}^4,
\end{equation}
where $g_s$ the spin-degeneracy factor of the $X^0$.

\section{Supernova simulation and EoS}

We use the latest 1D SN 1987A simulations including muons used to constrain the axion-muon coupling \cite{Bollig:2020xdr}. Four different simulations are performed changing the nuclear matter equation of state and the progenitor (and therefore the NS produced) mass.
These are labelled by SFHo-18.8, SFHo-18.6 and SFHo-20.0, or by LS220-20.0, depending on the EoS and mass of the progenitor star (in solar masses) used. The information given by the simulations are radial profiles of several thermodynamical quantities such as
density, particle abundancies or temperature at different times.

\begin{figure}[t]
    \centering
    \begin{subfigure}[t]{0.45\textwidth}
        \centering
        \includegraphics[width=\textwidth]{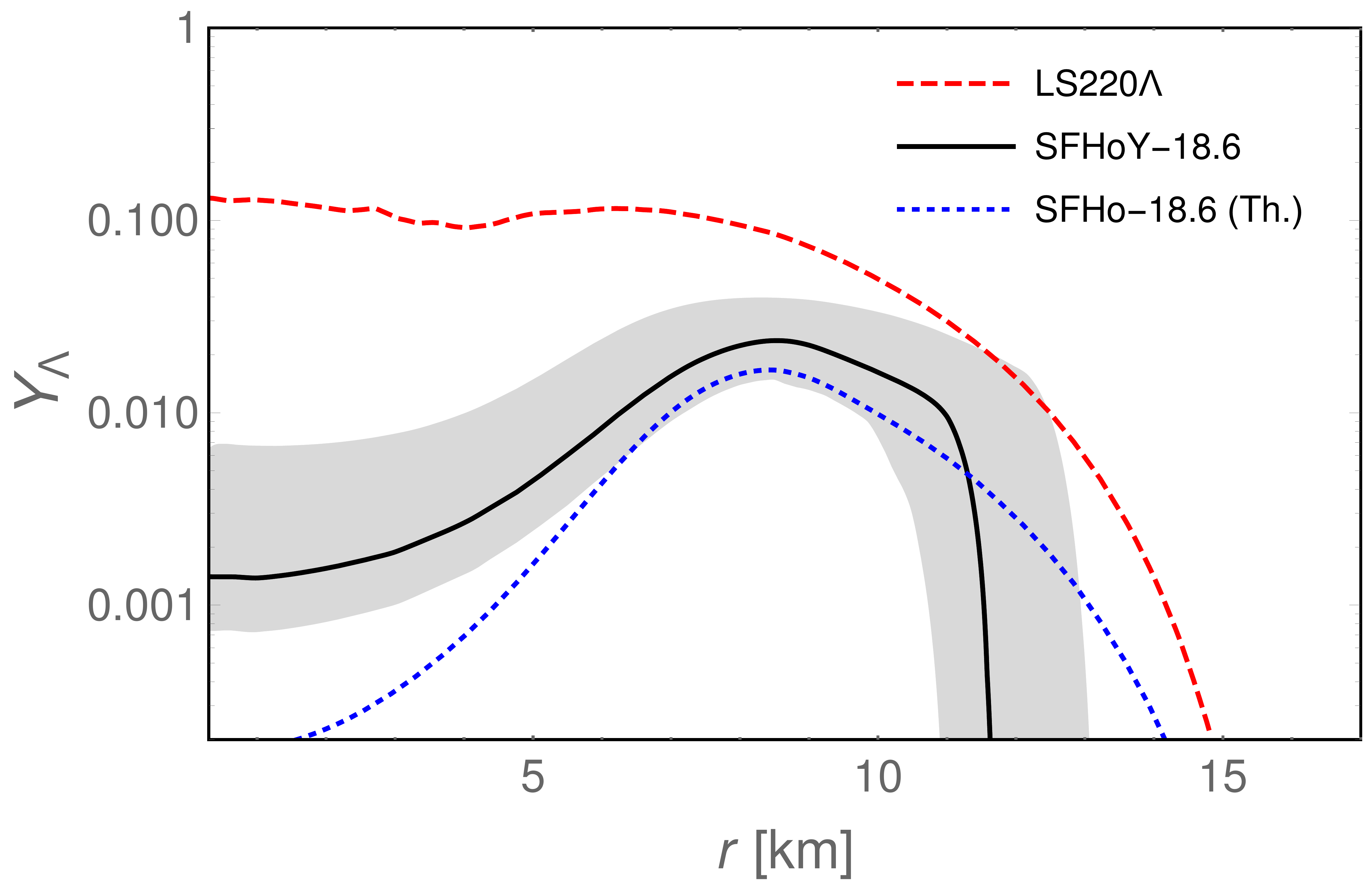} 
        \caption{\label{fig:Y_Lambda}} 
    \end{subfigure}
    \begin{subfigure}[t]{0.45\textwidth}
        \centering
        \includegraphics[width=\textwidth]{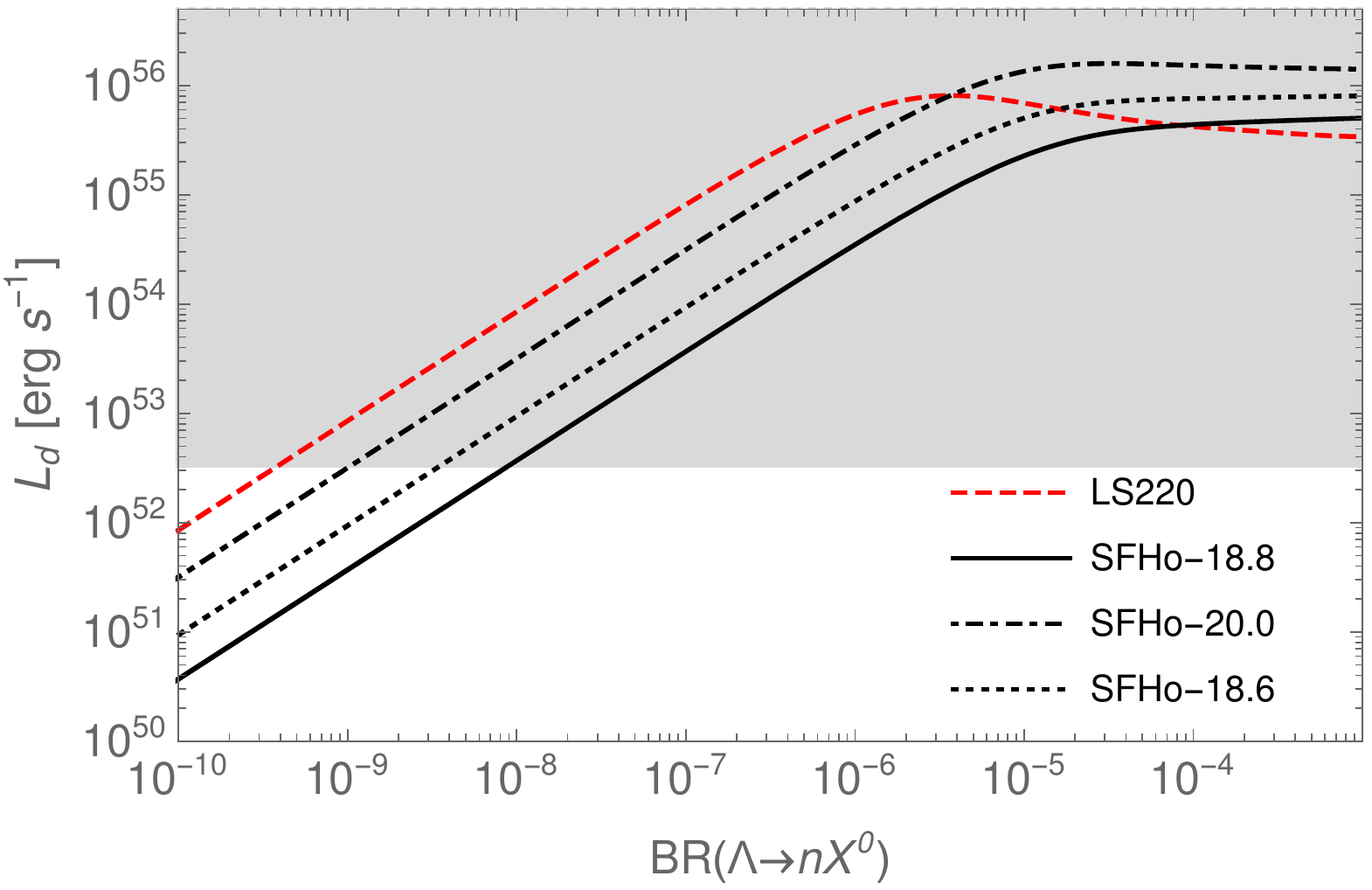} 
        \caption{ \label{fig:Results}} 
    \end{subfigure}
    \caption{(a) {\small Radial profiles of the $\Lambda$ abundancy $Y_\Lambda = n_\Lambda/n_B$, with $n_B$ the baryon number density using LS220 (red dashed) and SFHo EoS with hyperons. The gray band around the SFHoY-18.6 curve (solid black) correspond to the SFHo simulations with 20$M_\odot$ and 18.8$M_\odot$, respectively. We also include the results for SFHo-18.6 assuming a purely thermal distribution (blue dotted)
    .} (b) {\small Dark luminosity for the various simulations at $\sim1$ s post-bounce as a function of the branching-fraction of the decay $\Lambda\to n X^0$. Gray region is excluded by Eq.~\eqref{eq:boundRaffelt}.
    }}
\end{figure}

This simulations do not include hyperons, nonetheless, one can add them through the EoS extensions with hyperons: SFHoY \cite{Fortin:2017dsj} and LS220$\Lambda$ \cite{Oertel:2012qd,Gulminelli:2013qr}. The thermodynamical properties of the system remain unchanged (SFHoY) or they change only for extreme densities (LS220$\Lambda$) \cite{Fortin:2017dsj,Oertel:2016xsn} therefore 
we believe our approach is safe. Anyhow, LS220$\Lambda$ is not able to reproduce some neutron stars observations and will not be taken into account for our final result.
To compute the radial profiles of $\Lambda$ abundancies, shown in Fig. \ref{fig:Y_Lambda}, we make use of CompOSE database \cite{compose}.

Additionally, in \cite{Camalich:2020wac} we discuss how the medium effects affect in each EoS and how they are implemented in the simulations.

\section{Results}

With all of the computations described above we calculate the dark luminosity of SN 1987A as a function of $\Gamma(\Lambda\to n X^0)$. Now, using the bound of Eq. \ref{eq:boundRaffelt} we can set an upper limit on the branching fraction of that decay. In Fig. \ref{fig:Results} we show the luminosity as a function of the 
branching fraction for the different simulations. All in all, the SN 1987A bound is,

\begin{align}
    \label{eq:BRLimit}
    {\rm BR}(\Lambda\to n X^0)\lesssim8.0\times 10^{-9},     
\end{align}

obtained by combining the most refined calculation with the simulation giving the weaker bound (SFHo-18.80)~\cite{Bollig:2020xdr}. Note that this is a conservative limit because it stems from the simulation that produces the SN 1987A’s remnant on the low-mass edge of the allowed range and, therefore, have the coolest profile.

\subsection{Dark photons}

\begin{figure}[t]
    \centering
    \begin{subfigure}[t]{0.45\textwidth}
        \centering
        \includegraphics[width=\textwidth]{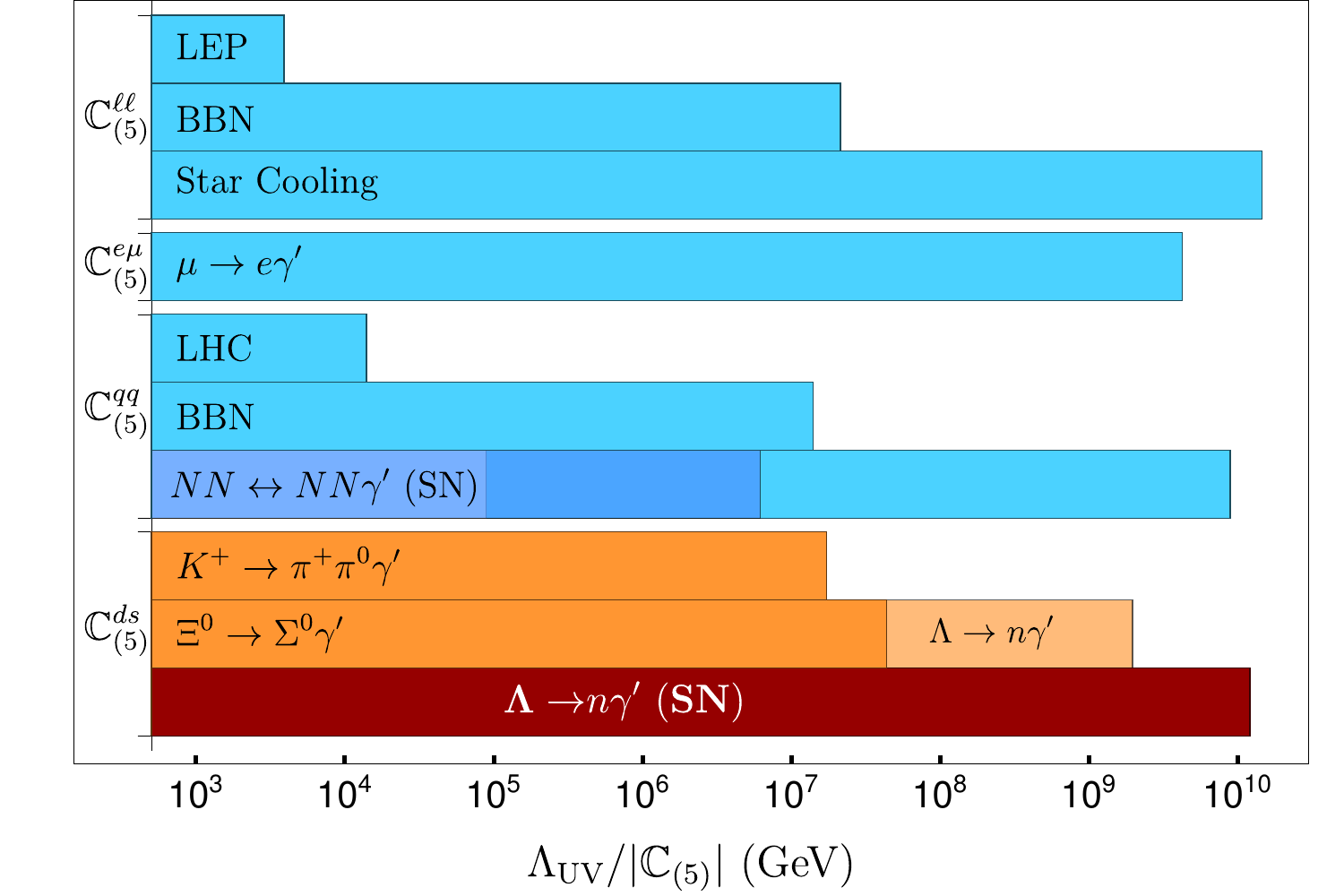} 
        \caption{\label{fig:DPlimits}} 
    \end{subfigure}
    \begin{subfigure}[t]{0.48\textwidth}
        \centering
        \includegraphics[width=\textwidth]{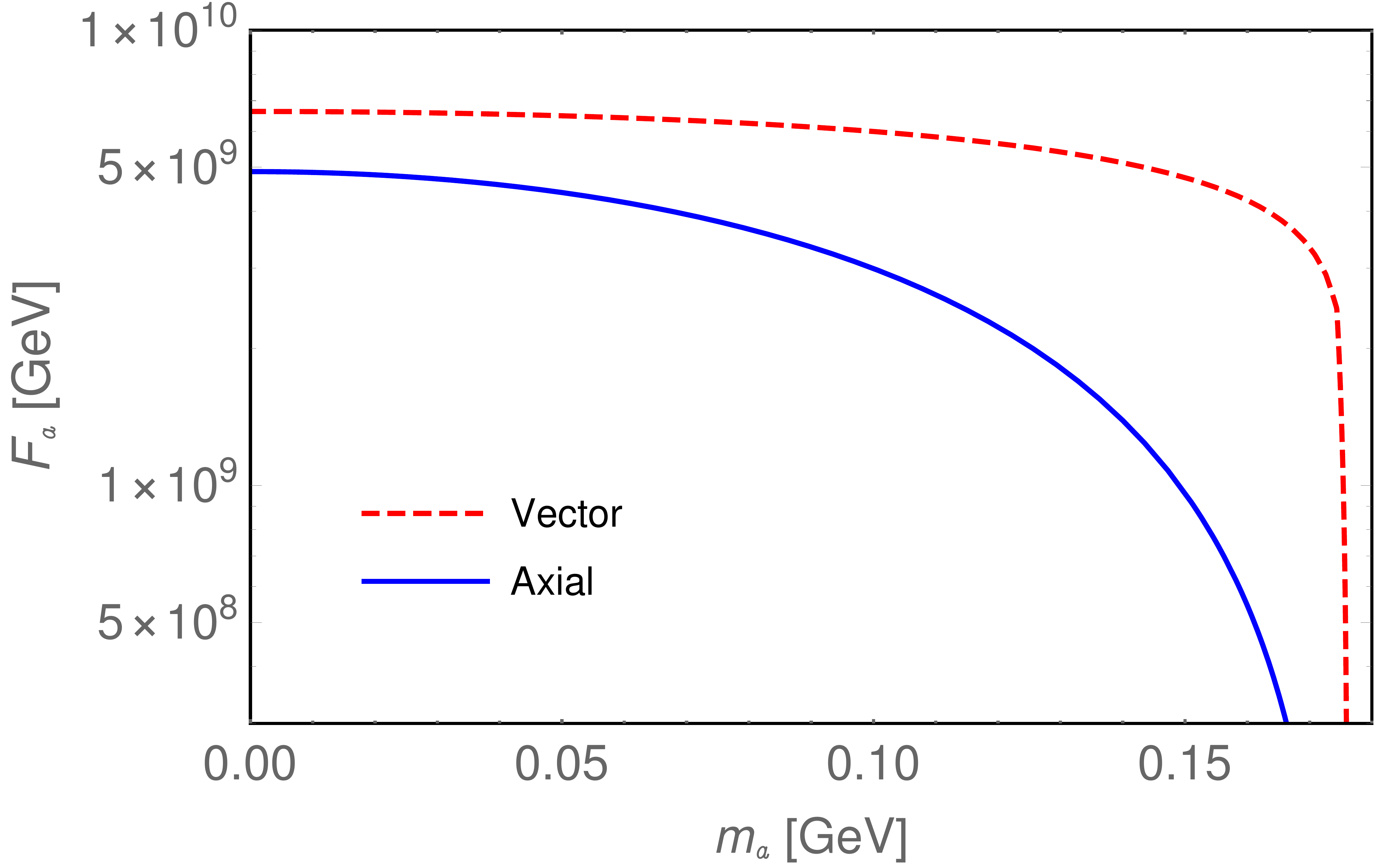} 
        \caption{ \label{fig:ALP}} 
    \end{subfigure}
    \caption{(a) {\small Model independent limits to the couplings of the dipole operator Eq.~(\ref{eq:LagrangianDP}) from various sources. In orange, limits to $ds$ coupling that can be directly compared to the one derived in this work, in dark red. In faint orange the prospected limit from BESIII, limits to other flavour configurations in blue.} (b) {\small Dependence  on the ALP mass of the lower bounds of $F_a = 2 f_a/c^{V,A}_{sd}$ for the vector or axial couplings (only one active at the time). The calculations have been done using the SFHo-18.8 simulation..
    }}
\end{figure}

We can apply our result to the massless dark photon case, if we consider the dimension-five operator
\begin{equation}
\label{eq:LagrangianDP}    
    \mathcal L_{\gamma^\prime}= \dfrac{1}{\Lambda_{\rm UV}}\bar{\psi}_i\sigma^{\mu\nu}\left(\mathbb{C}^{ij}+i \,  \mathbb{C}^{ij}_5\gamma_5\right)\psi_j F^{\prime}_{\mu\nu},
\end{equation} 
where $F^{\prime}_{\mu\nu}$ is the field strength associated to the dark photon, $\psi_i$ are the SM fermions and  $\mathbb{C}^{ij}_{(5)}$
are the couplings of the interaction, suppressed by the energy scale $\Lambda_{\rm UV}$, that depends on the underlying UV completion~\cite{Fabbrichesi:2020wbt}. Using Eq. \ref{eq:BRLimit} we can set the lower limit,
\begin{align}
\label{eq:LambdaUV}    
\Lambda_{\rm UV}\gtrsim1.2\times10^{10}\text{ GeV},
\end{align}
assuming $\sqrt{|\mathbb{C}^{ds}|^2+|\mathbb{C}^{ds}_5|^2}\sim 1$. In Fig. \ref{fig:DPlimits} we can see it compared to other limits on the same coupling coming from hyperon and kaon decays and to other quark and lepton couplings coming from cosmology, astrophysics and laboratories \cite{Fabbrichesi:2020wbt}.

\subsection{Axions}

The result is also valid for axions. Their couplings to SM fields are
\begin{align}
\label{eq:axion_couplings}
& \mathcal{L}_{a} = 
\frac{\partial_\mu a}{2 f_a} \, \bar \psi_i \gamma^\mu \big(c^V_{i j}+c^A_{ij} \gamma_5 \big) \psi_j \, , 
\end{align}
where $a$ is the axion field and $f_a$ is its decay constant. For the case of a massles axion, using the SN limit in Eq.~\eqref{eq:BRLimit} we obtain,
\begin{align}
\label{eq:Faxion}
F^V_{sd} \gtrsim7.1\times 10^9\text{ GeV,}~~~~F^A_{sd} \gtrsim5.2\times 10^9\text{ GeV},
\end{align}
for pure vector and axial couplings $F^{V,A}_{sd} \equiv 2 f_a/c^{V,A}_{sd}$, respectively.  In~\cite{MartinCamalich:2020dfe} this bound is compared exhaustively to others coming from astrophysics and laboratory searches. In particular, this constraint is also stronger than the SN bounds on the diagonal couplings to light quarks from nucleon-nucleon \textit{bremsstrahlung},  and on the leptonic couplings to $\mu\mu$ \cite{Bollig:2020xdr}.

In Fig. \ref{fig:ALP} we can see how these limits change with the presence of a massive ALP. In the appendix of \cite{Camalich:2020wac} it is discussed how the calculations change in presence of a massive dark boson instead of a massless one. It is interesting to note that the SN bound on massive ALPs can become comparable to the 
stringent bounds from laboratory experiments looking for $K^+ \to \pi^+ X^0$ in the 
two-pion decay region, where the sensitivity is strongly reduced due to the SM background.

\section{Conclusions}
The inclusion of $\Lambda$ hyperons in the proto-neutron star leads to new possible cooling mechanisms. We have used state-of-the-art simulations with the corresponding hyperonic EoS for our calculations to obtain the upper limit BR$(\Lambda\to nX^0)\lesssim8.0\times10^{-9}$. This is the strongest
bound that has been derived so far on the couplings of the massless dark photon to quarks. 
The analysis also puts strong constraints on flavor-violating axion models, and can be readily extended to other flavored dark sectors.

\section*{Acknowledgements}

This article is based on the talk given in 17th International Conference on Topics in Astroparticle and Underground Physics (TAUP 2021) \cite{youtube}. The original work \cite{Camalich:2020wac} was done in collaboration with Jorge Martin Camalich, Laura Tolós and Robert Ziegler. This work was supported by the Spanish Ministry of Science grants PGC2018-102016-A-I00, SEV-2015-0548 and FPI grant PRE2019-089992.

\section*{References}
\bibliographystyle{iopart-num}
\bibliography{Proceedings.bib}

\providecommand{\newblock}{}
\begin{thebibliography}{10}
\expandafter\ifx\csname url\endcsname\relax
  \def\url#1{{\tt #1}}\fi
\expandafter\ifx\csname urlprefix\endcsname\relax\def\urlprefix{URL }\fi
\providecommand{\eprint}[2][]{\url{#2}}

\bibitem{Bertone:2010zza}
Silk J {\em et~al.\/} 2010 {\em {Particle Dark Matter: Observations, Models and
  Searches}\/} (Cambridge: Cambridge Univ. Press) ISBN 978-1-107-65392-4

\bibitem{Bollig:2017lki}
Bollig R, Janka H~T, Lohs A, Martinez-Pinedo G, Horowitz C and Melson T 2017
  {\em Phys. Rev. Lett.\/} {\bf 119} 242702 (\textit{Preprint}
  \eprint{1706.04630})

\bibitem{Oertel:2016bki}
Oertel M, Hempel M, Kl\"ahn T and Typel S 2017 {\em Rev. Mod. Phys.\/} {\bf 89}
  015007 (\textit{Preprint} \eprint{1610.03361})

\bibitem{Bollig:2020xdr}
Bollig R, DeRocco W, Graham P~W and Janka H~T 2020 {\em Phys. Rev. Lett.\/}
  {\bf 125} 051104 (\textit{Preprint} \eprint{2005.07141})

\bibitem{Fortin:2017dsj}
Fortin M, Oertel M and Provid\^encia C 2018 {\em Publ. Astron. Soc. Austral.\/}
  {\bf 35} 44 (\textit{Preprint} \eprint{1711.09427})

\bibitem{Oertel:2012qd}
Oertel M, Fantina A and Novak J 2012 {\em Phys. Rev. C\/} {\bf 85} 055806
  (\textit{Preprint} \eprint{1202.2679})

\bibitem{Gulminelli:2013qr}
Gulminelli F, Raduta A, Oertel M and Margueron J 2013 {\em Phys. Rev. C\/} {\bf
  87} 055809 (\textit{Preprint} \eprint{1301.0390})

\bibitem{Oertel:2016xsn}
Oertel M, Gulminelli F, Provid\^encia C and Raduta A~R 2016 {\em Eur. Phys. J.
  A\/} {\bf 52} 50 (\textit{Preprint} \eprint{1601.00435})

\bibitem{compose}
CompStar Online Supernov$\ae$ -- Equations of State:\\
  \url{https://compose.obspm.fr/}

\bibitem{Camalich:2020wac}
Martin~Camalich J, Terol-Calvo J, Tolos L and Ziegler R 2021 {\em Phys. Rev.
  D\/} {\bf 103} L121301 (\textit{Preprint} \eprint{2012.11632})

\bibitem{Fabbrichesi:2020wbt}
Fabbrichesi M, Gabrielli E and Lanfranchi G 2020  (\textit{Preprint}
  \eprint{2005.01515})

\bibitem{MartinCamalich:2020dfe}
Martin~Camalich J, Pospelov M, Vuong P~N~H, Ziegler R and Zupan J 2020 {\em
  Phys. Rev. D\/} {\bf 102} 015023 (\textit{Preprint} \eprint{2002.04623})

\bibitem{youtube}
Talk recorded and available at:\\
  \url{https://www.youtube.com/watch?v=lZ5eb8888uA}

\end{thebibliography}

\end{document}